\documentstyle[aps,prb,multicol]{revtex}
\begin{document}
\newcommand{\sia}{$\partial\sigma/\partial a$}
\newcommand{\sic}{$\partial\sigma/\partial c$}
\newcommand{\siu}{$\partial\sigma/\partial u$}
\newcommand{\kvec}{{\bf k}}
\newcommand{\vv}{\!\!\!\!\!\!}
\newcommand{\vareps}{\varepsilon}  
\newcommand{\vare}{\varepsilon}  
\newcommand{\eps}{\epsilon}  
\newcommand{\De}{$\Delta$}
\newcommand{\de}{$\delta$}
\newcommand{\mc}{\multicolumn}
\newcommand{\be}{\begin{eqnarray}}
\newcommand{\ee}{\end{eqnarray}}
\newcommand{\Pvec}{{\bf P}}

\draft \title{Spontaneous polarization
and piezoelectric constants 
of III-V nitrides}
\author{Fabio Bernardini and Vincenzo Fiorentini}

\address{INFM -- Dipartimento di Scienze Fisiche, Universit\`a di 
Cagliari, I-09124 Cagliari, Italy}

\author{David Vanderbilt}
\address{ Department of Physics and Astronomy, Rutgers University, 
Piscataway, NJ, U.S.A.}

\maketitle

\begin{abstract}
The 
spontaneous  polarization, dynamical Born charges, and  piezoelectric
constants of the III-V nitrides AlN, GaN, and InN 
are studied {\it ab initio} using the Berry phase approach to
polarization in solids. The piezoelectric constants
   are found to be  up  10 times
larger than in conventional III-V's and II-VI's, and comparable 
to those of ZnO. Further properties
 at variance with those of conventional  III-V compounds are the
 sign of the piezoelectric constants (positive as in II-VI's)
 and the very large spontaneous polarization.
\end{abstract}

\begin{multicols}{2}

In this paper we report an ab initio study of 
the
 spontaneous polarization,
 piezoelectric constants, and dynamical charges 
of the III-V nitride semiconductors AlN, GaN, and InN.
\cite{nature} 
This class of polarization-related  properties is of obvious importance for the
study of nitride-based piezodevices \cite{khan} and 
  multilayer structures. In
 particular, the knowledge of these properties allows an
insightful treatment
 of the polarization (and ensuing electric fields)
in strained and polarized nitride junctions and superlattices
 under any strain  condition, as discussed elsewhere.\cite{interf}
From the  present study, one of the first applications of the modern
theory of polarization 
 in solids \cite{KS.PRB47,restan} 
to real and ``difficult'' materials
of technological interest (the first to our knowledge using
pseudopotentials),  the nitrides emerge as highly unusual III-V materials, 
resembling II-VI oxides and in some respects ferroelectric perovskites. 
The results we report here are of special  interest in view of
the scarcity of the data  (both experimental and  theoretical) available
at present for the nitrides.

In the absence of external fields,
the total macroscopic polarization {\bf P} of a solid
  is the sum of the
spontaneous polarization  ${\bf P}^{\rm eq}$ in the equilibrium
structure, and of the strain-induced or piezoelectric polarization
$\delta {\bf P}$.  
In the linear regime, the piezoelectric   polarization
is  related  to the strain $\eps$ by 
\be
\delta P_i = \sum_j e_{ij} \eps_j 
\ee
which defines  
 the components of the 
piezoelectric tensor $e_{ij}$ (Voigt notation is used).

The  natural structure of 
the III-V nitrides is wurtzite,
 an hexagonal crystal structure defined by  the 
edge length $a$ of the basal hexagon, the height $c$ of the  hexagonal
prism, and an internal parameter $u$ defined as the
anion-cation bond length  along the  (0001) axis in units of $c$.
Wurtzite is the  structure  with highest symmetry compatible with
the existence of spontaneous polarization, 
\cite{nye,post} and  the piezoelectric tensor of wurtzite has three 
non-vanishing independent components.
Therefore, the polarization in
these materials system will both  a spontaneous  and a piezoelectric
component.

In this work we shall restrict ourselves to  polarizations along the 
(0001) axis;  this is the direction along which both standard bulk
materials and nitride superlattices are grown.
 The spontaneous polarization
vector then points in the direction of the $c$
axis, ${\bf P}^{\rm eq} = P^{\rm eq}\, \hat{\bf z}$, and
the piezoelectric  polarization  is
simply expressed via the piezoelectric coefficients $e_{33}$ and 
$e_{13}$  as 
\be 
\label{eq.piezo}
\delta P_3 = e_{33}\eps_3 + e_{31}(\eps_1 + \eps_2), 
\ee 
where $\eps_3 = (c - c_0)/c_0$  is the strain along the $c$ axis,
$\eps_{1} = \eps_{2} = (a - a_0)/a_0$ is the strain in the basal plane, 
 $a_0$ and $c_0$ being the equilibrium values
of the lattice parameters.
The third independent component of the piezoelectric 
tensor, $e_{15}$,  is related to the polarization induced by a shear strain,
and will not be considered in this work. 

To linear order, the change in polarization 
can also be expressed as
\be
\label{eq.pol}
\delta P_3 = \frac{\partial P_3}{\partial a} (a - a_0) 
	   + \frac{\partial P_3}{\partial c} (c - c_0) 
	   + \frac{\partial P_3}{\partial u} (u - u_0, 
\ee  
where $a_0$, c$_0$, and $u_0$ are the equilibrium lattice constants and
internal parameter. (Note that the parameters $a$, $c$ and $u$ are not
independent, because of the 
condition of vanishing forces  along the axis direction.) 
Once  the polarization derivatives in Eq.~\ref{eq.pol}
are known, the  macroscopic 
piezoelectric tensor coefficients can  be expressed as
\be
\label{eq.e33}
e_{33} &=&\,\, c_0\, \frac{\partial P_3}{\partial c   }  
       + \frac{    4e c_0  }{\sqrt{3}a_0^2} \,Z^*\, 
         \frac{d u}{d c }  
\ee
and
\be
\label{eq.e31}
e_{31} &=& \frac{a_0}{2}\, \frac{\partial P_3}{\partial a   }  
       + \frac{    2e      }{\sqrt{3}a_0} \,Z^*\,
         \frac{d u}{d a},  
\ee 
where 
\be
\label{eq.Z}
Z^* = \frac{\sqrt{3}a_0^2}{4e} \frac{\partial P_3}{\partial u} \equiv
{\cal Z}_3^{(T)}
\ee 
is the axial component of the  Born, or transverse,
dynamical charge tensor ${\cal Z}^{(T)}$.
In the last equation above, it is implicit that 
the vector  connecting the cation with the anion
 has modulus  $uc$ and  points in the direction of the $c$ axis.
The first term in Eqs. \ref{eq.e33} and \ref{eq.e31} is called 
the clamped-ion term, and it represents the effect of the strain on
the electronic structure. The second term quantifies the effects of
{\it internal} strain  on the polarization.
The
 derivatives of $u$ with respect to $c$ and $a$ in Eqs. 
\ref{eq.e33} and \ref{eq.e31} are related to the
strain derivatives of $u$  through
$
c_0\, {d u}/{d c} =  {d u}/{d \eps_3}$ and
$a_0\, {d u}/{d a} = 2\, {d u}/{d \eps_1}.$

We have calculated the polarization within the Berry phase approach of 
Ref. \onlinecite{KS.PRB47}. The method's 
output is the difference in polarization among two
states of a system, provided they can be connected 
 by an 
  adiabatic transformation which leaves the system insulating.
 The difference  of electronic polarization $\bf \Delta P_e$
among two systems is then related to the geometric quantum phase by 
\be 
\label{eq.berry}
{\bf \Delta P_e} 
	      =\Pvec_e(\lambda_2) - \Pvec_e(\lambda_1) 
\ee
where 
\be
\Pvec_e(\lambda)= 
- \frac{2e}{(2\pi)^3} 
\left.\int_{BZ} d\,\kvec \frac{\partial }{\partial \kvec'}\,\,
	     \phi^{(\lambda)}({\bf k,\,k'})\right\vert_{\bf k'=k},
\ee
 the integration domain in momentum space is 
the reciprocal unit cell,
 $\lambda$  characterizes the adiabatic
transformation,
and  $\phi^{(\lambda)}$ is the geometric quantum phase
\be
\phi^{(\lambda)}({\bf k,\,k'}) =
     {\rm Im}\, \{\ln\, [\det 
     \langle u_m^{(\lambda)}({\bf k})\vert\, u_n^{(\lambda)}({\bf k'})
     \rangle] \label{bphase}
\ee
of the occupied crystal Bloch states
$u_n^{(\lambda)}({\bf k})$.
The definition of the geometric quantum phase 
introduces an arbitrary constant in the value of the polarization 
so that the latter is well-defined only modulo $e{\bf R}/\Omega$,
with ${\bf R}$  a real-space lattice vector.
This is not a difficulty in practice, as 
one is generally interested in
 polarization changes such that
$\vert {\bf \Delta} \Pvec \vert \ll \vert e{\bf R}/\Omega \vert$,
so that no ambiguity can arise.

The adiabatic transformation labeled by $\lambda$ is trivially
identified as far as the strains $\eps_1$, $\eps_2$, and  $\eps_3$  
are
concerned, so that the strain derivatives of the polarization within
the wurtzite 
structure are easily calculated. On the other hand,  
the {\it
absolute} value of the spontaneous polarization of the 
 wurtzite structure (state $\lambda_2$ in Eq. \ref{eq.berry}) 
must  be calculated as a difference relative to a
structure having zero polarization, and functioning as reference state
$\lambda_1$, since  only
polarization differences can be meaningfully calculated within the
modern theory,\cite{KS.PRB47,restan}: zincblende is the natural choice
for this purpose.  One needs not actually
devise a gap-preserving adiabatic transformation 
turning zincblende into wurtzite (a quite non-trivial task):
 the interface theorem of Ref. \onlinecite{Vand.PRB48}
proves that  Eq. \ref{eq.berry}  can be used directly
 provided that an insulating interface can be built connecting the two
different structures. As this is found to be  the case \cite{interf}
 for the systems of relevance
here, the absolute polarization of wurtzite nitrides can be  easily
computed 
(see also Ref. \onlinecite{post}).
 
All the present  first principles calculations are based on density 
functional theory \cite{dft} in the local density approximation 
(LDA) for the exchange-correlation energy functional,
for which we adopt the 
 Ceperley-Alder \cite{CA} form as parametrized by Perdew and 
Zunger.\cite{PZ}
The wave functions are expanded in a plane-wave basis set up to an energy
cutoff of 25 Ry.
Ultrasoft pseudopotentials 
\cite{USPP} have been employed for all the elements 
involved in the calculations. In particular,   
the Ga and In pseudopotentials are constructed so as to 
explicitly include  the semicore {\it d} electrons in the 
valence, resulting in a very accurate description of the electronic
and structural properties. 

All calculations for the wurtzite structure
used a mesh of 12 Chadi-Cohen special points
for the Brillouin zone sampling.\cite{cc}
 The k-space integration   for computing the polarization was instead
done on a (4,4,10) Monkhorst-Pack \cite{mp} mesh; this amounts to 16 
 $\kvec_{\perp}$ points times  10 $\kvec_{\|}$  points. Refining  the
mesh to  360 points (the  (6,6,10) mesh) gave no appreciable variation
of the calculated values.
The partial derivatives of the polarization  appearing 
in Eq.~\ref{eq.pol} are 
 easily evaluated numerically 
through the  polarization change  induced by 
typically a $\pm$2\% change in the crystal  parameters $a$ and $c$
(with $u$ kept fixed, i.e. clamped ion) and in the internal parameter
$u$, around their equilibrium values.
 
\par
 Since  we are interested in the spontaneous polarization, 
a careful determination of the equilibrium 
structure parameters is needed.
The three independent structural  parameters 
 have been obtained by  polynomial interpolation of 
 the  total energy values calculated over a grid 
of values of $a$ and $c$. The internal parameter $u$ was
optimized  at each ($a$,$c$) grid point following  the
Hellmann-Feynman  forces. The results,
summarized in Table \ref{tab.struc}, are very close to 
those reported in our previous work.\cite{Satta}

In Table~\ref{tab.pol} we report the  spontaneous polarization,
dynamical charges, piezoelectric coefficients (and parts thereof) for 
the III-V nitrides. Also included are the
same quantities calculated  for
ZnO and BeO (with the Zn 3$d$ states included in the valence).
In addition, we list the same quantities as calculated
in Ref.~\onlinecite{dalcorso.zno},
 using the  Berry phase approach and the FLAPW method,
 for
ZnO and BeO.
 The agreement of our results with experiments 
for the latter materials is quite good.
To facilitate further comparison with other III-V and II-VI
 systems, we have collected 
 in Table \ref{tab.pol2} the 
 piezoelectric constants for a number of III-V and II-VI
compounds from Refs. \onlinecite{Gironcoli.PRL},
\onlinecite{DalCorso.CdTe}, and
\onlinecite{Gironcoli.Ferroel}, 
calculated by density functional linear response theory.\cite{bgt}

To compare zincblende and wurtzite structures in Tables \ref{tab.pol}
and \ref{tab.pol2},
we have transformed the zincblende piezoelectric tensor by
rotating to a coordinate system with $z$ along a (111) direction.  
Thus,
$e_{33} = 2 \,e_{14} /\sqrt{3}$ and  $e_{31} =
 -e_{14}/\sqrt{3},$
with $e_{14}$  the piezoelectric constant of the zincblende
material referred to the cubic axes.
This direct comparison is meaningful since
(1) the possible deviations from the ideal wurtzite
structure do not influence significantly the polarization {\it derivatives},
and (2) it is known that in polytypical materials the values 
of the piezoelectric constants 
in the two competing structures 
agree  to within a
few percent (modulo the appropriate
coordinate transformation). \cite{dalcorso.zno}

The III-V  nitrides exhibit rather striking and
unusual   properties. Overall they resemble II-VI oxides,
and ZnO in particular, and appear to be very different 
from conventional   III-V semiconductors.
Several features of the data  in Tables \ref{tab.pol} and
\ref{tab.pol2} confirm this statement:

\noindent
1) The absolute value of the piezoelectric constants is up to 
about 10 times larger than in conventional III-V's and II-VI's. In 
particular, both the constants 
${e}_{33}$ and  ${e}_{31}$ of AlN are larger than those of ZnO,
and are therefore the largest known so far among the tetrahedrally
bonded semiconductors.

\noindent
2) The spontaneous polarization, i.e. the polarization at zero strain,
is very large in the nitrides. That of AlN, in particular, is the largest
of our set of semiconductors, and only about 3-5 times smaller than
in typical ferroelectric perovskites.\cite{zhong}
Indeed, the piezoelectric constants themselves are
 only about 3 times smaller than in those 
ferroelectrics.\cite{rabe} 

\noindent
3) The piezoelectric constants have the same sign as in II-VI
 compounds, and opposite to 
III-V's. While in normal III-V's the
 clamped-ion term 
is larger in absolute value than the  internal-strain ionic
contribution, 
in the nitrides the latter prevails due to the larger $Z^*$.
Compared to normal III-V's,
this sign inversion is a qualitative difference 
of  obvious practical relevance.
\cite{nota}

\noindent 
4) The nitrides follow qualitatively a well defined III-V trend: the
piezoelectric constants increase in magnitude  as a function of the
anion chemical identity 
 as one moves upwards within Period V, i.e. from Sb to N, because
the  ionic contribution tends to prevail over the  electronic
``clamped-ion'' term as the  anion becomes lighter
(note  that a similar trend is also obeyed by Zn II-VI's).
The nitrides are an extreme case of this trend, and
 their piezoelectric response 
is by far larger than that of all other III-V's
(see Table \ref{tab.pol2}), and of opposite sign.

\noindent
5) The Born
effective charge in the nitrides
 is within 10\% of the nominal ionic charge,
i.e. the ratio  $Z^*/Z_{\rm ion}$ is about unity. 
The same ratio is at most  0.7 in III-V compounds.
In this context, it is appropriate  to mention that the
 effective charges have no rigorous  relation to
nominal ionicity (although they may be  a reasonable 
qualitative measure of ionicity in many binary compounds, suggesting
trends analogous to those given by e.g. the charge asymmetry $g$
parameter of Garcia and Cohen.\cite{garcia})
In particular, $Z^*$ needs not be smaller than or
equal to the nominal ionic charge except in a purely ionic picture;
conversely, whereas $Z^* < Z_{\rm ion}$ may be interpreted 
as ``smaller ionicity'' in some sense, $Z^* > Z_{\rm ion}$ is not
meaningful in an ionic picture. Effective charges much larger than  
the nominal ionic charge are common in perovskites\cite{zhong}
and (to a lesser extent) even in  alkaline earths oxides.\cite{resta2} 
This behavior is due  to covalency and correlation contributions.
We find (see Table \ref{tab.pol})  $Z^* > Z_{\rm ion}$ 
for InN, but with a deviation of $\sim$1\%, i.e. the effective charge
is  again very close to nominal ionicity.

Because of the sensitive dependence of the
polarization on the structural parameters, 
there are some quantitative differences in spontaneous polarization
for the three nitrides. For instance, the 
increasing non-ideality 
of the structure  going from GaN to InN to AlN
($u$ gets longer and $c/a$ gets shorter), 
corresponds to an increase of the spontaneous polarization.
To check this, we calculated the spontaneous polarization of AlN
using the ideal structural parameters $u_{\rm id}=0.375$ and
$c_{\rm id}=\sqrt{8/3}\, a$, and obtained a
spontaneous polarization equal to 
--0.033 C/m$^2$, very similar to GaN and InN, and a factor of 2.5
smaller than the spontaneous polarization in the actual equilibrium
structure. The piezoelectric constants (i.e. polarization derivatives)
are instead expected to vary mildly.

Additional III-V--like trends to be observed are that in III-V's the effective charge
has similar values for AlAs and GaAs, but a rather larger 
value for InAs.\cite{Gironcoli.PRL} This
trend is clearly obeyed by the nitrides  as well, indeed in a more
clearcut way than in most other  III-V's.
Another interesting  trend  is the connection
between the chemical nature of the cation and the piezoelectric constant.
 The piezoelectric constant for Al and In III-V's
was found to be  \cite{Gironcoli.PRL}
 consistently larger than the corresponding value for Ga compounds.
This is also true of the nitrides, where the deviations are
again larger. Since our calculations include the
semicore  $d$ states in the valence for both In and Ga, 
this trend is   not an artifact of the  neglect of semicore
states as hypothesized in Ref. \onlinecite{Gironcoli.PRL}.

While it is known that treating
 the semicore $d$ states as valence is necessary
to obtain good structural properties for GaN and InN,\cite{Satta,fiore}
 neither the spontaneous polarization nor
 the effective charges are influenced strongly. 
Using a Ga pseudopotential in which the 3$d$ states had been
frozen in the core, we recalculated the spontaneous polarization
 and effective charges of GaN using the structural parameters
 given in Table \ref{tab.struc}: 
the results are $P^{\rm eq} = -0.031$ C/m$^2$ and $Z^* = 2.73$,
showing that the direct influence of the 3$d$ states on the final
polarization value is  minor. This was to be expected since
 the $d$ bands in GaN are quite flat
over most of the Brillouin zone, so that the overlap matrix 
(Eq. \ref{bphase}) is nearly the identity for those states, and
its contribution to the Berry phase is small.

Experimental data to compare our predictions with are very scarce.
 No data is available for the spontaneous polarization,
 a quite elusive quantity in this respect; our
calculated values will be a useful input for device modeling.
\cite{khan,interf}
Also in the case of  the piezoelectric constants, our predicted
data should be of use to future experimental work. The
only experimental data  available\cite{ieee} 
to our knowledge  are for AlN,
${e}_{33}^{\rm exp} = 1.55$ C/m$^2$, and
${e}_{31}^{\rm exp} = -0.58$ C/m$^2$. Our 
predicted values of 1.46 C/m$^2$ and --0.60 C/m$^2$ 
agree with experiment to within $\sim 5$\%.

\par
In summary, we have calculated the spontaneous polarization,
dynamical charges, and piezoelectric  constants of III-V nitrides
 by means of the Berry phase approach.
The results show that III-V nitrides resemble 
 II-VI compounds in terms of the sign
of  the spontaneous polarization and of
the piezoelectric constants; the latter  constants are much larger
in absolute value than those of all III-V's and II-VI's,
except for ZnO. The values of
spontaneous polarization are comparable  to
or larger than those of II-VI's;
 AlN has the largest value reported so far for any binary
compound, and only a factor of 3 to 4 smaller than in typical perovskites.
Some remnant of normal III-V--like behavior survives, embodied in  the
 trends of the piezoelectric constants and effective charges 
when the  chemical identity of the anion or cation changes.

V.F and F.B. acknowledge partial support by the European Community
through Contract BRE2-CT93, and by CINECA Bologna through
Supercomputing  Grants. D.V. acknowledges support of 
ONR Grant N00014-97-1-0048.

\narrowtext
\begin{table}[ht]
\caption{Structural parameters for AlN, GaN and InN.} 
\begin{tabular}{lccc}
       & $a_0$ (bohr)&$c_0/a_0$ &$u_0$   \\ \hline 
AlN    & 5.814 & 1.6190& 0.380   \\
GaN    & 6.040 & 1.6336& 0.376   \\
InN    & 6.660 & 1.6270& 0.377   
\end{tabular}
\label{tab.struc}
\end{table}

\begin{table}[ht]
\caption{Calculated   spontaneous polarization (in units of C/m$^2$), 
Born effective charges (in units of $e$), and piezoelectric constants
 (in units of C/m$^2$)
for III-V wurtzite nitrides and
II-VI wurtzite oxides.}
\begin{tabular}{lddddddd}
       &$P^{\rm eq}$ & $Z^*$& $du/d\epsilon_3$&${e}_{33}$
&${e}_{31}$ &${e}_{33}^{(0)}$&${e}_{31}^{(0)}$  \\
\hline
AlN  &--0.081 &--2.70  &--0.18 &  1.46   & --0.60  & --0.47 & 0.36 \\
GaN  &--0.029 &--2.72  &--0.16 &  0.73   & --0.49  & --0.84 & 0.45 \\
InN  &--0.032 &--3.02  &--0.20 &  0.97   & --0.57  & --0.88 & 0.45 \\
ZnO  &--0.057 &--2.11  &--0.21 &  0.89   & --0.51  & --0.66 & 0.38 \\
BeO  &--0.045 &--1.85  &--0.06 &--0.02   & --0.02  & --0.60 & 0.35 \\ 
\hline
ZnO$^a$  &--0.05  &--2.05 &--0.25 &  1.21 & --0.51&--0.58 & 0.37 \\
BeO$^a$  &--0.05  &--1.72 &--0.09 &  0.50 & ---   &--0.29 & ---  \\ 
\end{tabular}
$^a$ Reference \onlinecite{dalcorso.zno}
\label{tab.pol}
\end{table}

\begin{table}[ht]
\caption{Piezoelectric constants (in units of C/m$^2$) 
of several zincblende compounds,
calculated via linear response theory in
Refs. \protect\onlinecite{Gironcoli.PRL} (III-V), 
 \protect\onlinecite{DalCorso.CdTe} (CdTe) and
 \protect\onlinecite{Gironcoli.Ferroel} (other II-VI).}
  \begin{tabular}{lddldd}
         & $e_{33}$ & $e_{31}$ &
         & $e_{33}$ & $e_{31}$ \\ \hline 
CdTe     &  0.03 &--0.01  & 
AlAs     &--0.01 &  0.01  \\
ZnS      &  0.10 &--0.05  & 
GaAs     &--0.12 &  0.06  \\
ZnSe     &  0.04 &--0.02  & 
InAs     &--0.03 &  0.01  \\
AlP      &  0.04 &--0.02  & 
AlSb     &--0.04 &  0.02  \\
GaP      &--0.07 &  0.03  & 
GaSb     &--0.12 &  0.06  \\
InP      &  0.04 &--0.02  & 
InSb     &--0.06 &  0.03  
\end{tabular}
\label{tab.pol2}
\end{table}

\end{multicols}

\begin{thebibliography}{99}
\bibitem{nota}
 AlP and InP have indeed \cite{Gironcoli.PRL} 
a  positive, but very small, $e_{14}$, which is again related to
larger-than-average $Z^*$.
\end{thebibliography}
\end{document}